\documentclass[%
 aip,
 amsmath,amssymb,
 reprint,%
]{revtex4-1}

\usepackage{graphicx}
\usepackage{dcolumn}
\usepackage{bm}

\usepackage[utf8]{inputenc}
\usepackage[T1]{fontenc}
\usepackage{mathptmx}
\usepackage{etoolbox}

\usepackage{siunitx}
\usepackage{tikz}
\usepackage{pgfplots}
\usepgfplotslibrary{groupplots}
\usetikzlibrary{spy,decorations.fractals,shapes.misc}
\usetikzlibrary{decorations.markings}
\usepgfplotslibrary{colormaps}
\usepgfplotslibrary{colorbrewer}
\usetikzlibrary{arrows.meta}
\pgfplotsset{cycle list/Dark2}
\pgfplotsset{compat=1.18}
\usepackage{subcaption}

\usepgfplotslibrary{external}
\tikzset{external/system call={pdflatex \tikzexternalcheckshellescape -halt-on-error
		-interaction=batchmode -jobname "\image" "\texsource" &&
		pdftops -eps "\image.pdf"}}

\makeatletter
\def\@email#1#2{%
 \endgroup
 \patchcmd{\titleblock@produce}
  {\frontmatter@RRAPformat}
  {\frontmatter@RRAPformat{\produce@RRAP{*#1\href{mailto:#2}{#2}}}\frontmatter@RRAPformat}
  {}{}
}%
\makeatother
\begin{document}

\preprint{AIP/123-QED}

\title[A HCAO approach for polyatomic chemistry modeling in DSMC]{A harmonically-coupled-anharmonic-oscillator approach for polyatomic chemistry modeling in DSMC}
\author{S. Lauterbach}
\author{F. Tuttas}%
 \email{tuttasf@irs.uni-stuttgart.de}
\author{M. Pfeiffer}
\affiliation{
Institute of Space Systems, University of Stuttgart, Pfaffenwaldring 29, 70569 Stuttgart, Germany
}%

\date{\today}

\begin{abstract}
Atmospheric entry processes are characterized by high-enthalpy gas flows in strong thermo-chemical non-equilibrium.
Accurate simulations of such conditions remain challenging due to the extreme conditions and the complex influence of internal energy modes.
In particular, the common assumption of uncoupled harmonic vibrations may break down, and excited internal energy states can directly influence reaction rates.
Previously, an anharmonic oscillator model has been developed by Civrais et al.~\cite{clement} to improve the accuracy of the Direct Simulation Monte Carlo (DSMC) method under such conditions.
However, this extension has so far been limited to diatomic molecules.

To increase the accuracy of the DSMC method in the open-source code PICLas, the anharmonic oscillator model is extended to include polyatomic species.
The proposed model explicitly considers anharmonic effects and intramolecular energy redistribution.
Vibrational degrees of freedom are treated in a local mode basis, in which anharmonic stretching modes are harmonically coupled by harmonic bending modes.
The coupling allows for the redistribution of localized vibrational excitation.
Dissociation can occur by the strong excitation of a stretching mode, and specific modes can be coupled to the reaction coordinate of the transition state in bimolecular exchange reactions.

The newly developed model is evaluated by the comparison to high-fidelity calculations for a set of representative processes.
Investigated are different dissociation reactions, which exhibit a high degree of energy redistribution, and the hydrogen-exchange reaction between methane and a hydrogen radical, in which only selected modes contribute to the reactive process.
In addition, the recombination-dissociation equilibrium system has been investigated for methane.
\end{abstract}

\maketitle


\section{Introduction}
\begin{figure}
   \centering







    \includegraphics{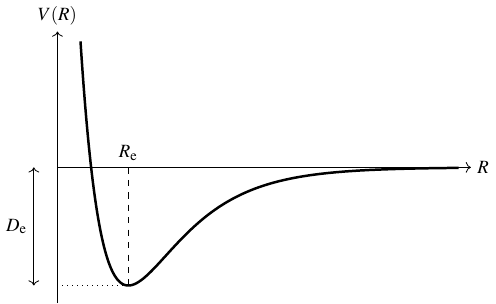}
    \caption{Potential energy curve for a diatomic molecule in dependence of the bond length. Additionally, the equilibrium bond distance $R_\text{e}$ is shown, as well as the depth of the potential, corresponding to the dissociation energy $D_\text{e}$.} \label{fig:PES}
\end{figure}

Molecular vibrations play a central role in the determination of energy transfer processes and chemical reaction kinetics.
The basis for the description of molecular vibration and reactive cross-sections is the multi-dimensional Potential Energy Surface (PES), which represents the effective potential energy of a system as a function of all nuclear coordinates.
Within the Born--Oppenheimer approximation, electronic and nuclear motions are separated, so that the PES acts as an effective potential governing nuclear dynamics.
For a diatomic system, the PES only depends on the internuclear distance, while for polyatomic systems the dimensionality increases substantially.
An examplatory potential function for a diatomic system is shown in Figure~\ref{fig:PES}.
Here, the potential function only shows a dependence on the bond length. The potential energy shows a minimum at the equilibrium bond distance, with a well depth corresponding to the dissociation energy, and increases fast for lower distances. As the bond length increases, the energy increases accordingly, until a value of zero is reached, as the length approaches infinity. Important properties of the molecular system can be directly associated with characteristic features of the PES.
Minima on the PES correspond to equilibrium molecular structures, with the global minimum corresponding to the most stable configuration of the system.
The vibrational properties and spectra are related to the curvature of the surface close to a minimum.
Between two minima on a PES, a minimum energy path through a first-order saddle point, corresponding to the transition state, can be found along which a reaction takes place.
The relative potential height defines the activation energy of a given reaction.
Due to these relations with physical quantities, the identification of stationary points is a central objective in computational chemistry.
Commonly, the hypersurface close to stationary points in these calculations is approximated by a second-order Taylor expansion, which leads to the harmonic oscillator model on the basis of normal mode analysis.
Within this approximation, the quantized energy eigenvalues are equally spaced and given by:
\begin{equation}
    E_\text{vib}^\text{HO}(i) = h \nu_0 \left( i + \frac{1}{2} \right), \quad i=0,1,2,\dots \label{eq:energy-SHO}
\end{equation}
In the above equation, $i$ is the vibrational quantum number, $h$ is the Planck constant, and $\nu_0$ is the fundamental frequency, which is related to the reduced mass and the bond strength of the molecule.
The quantum harmonic oscillator has a non-zero energy value in the ground state, as characterized by the zero-point energy.
For low-lying vibrational states, the harmonic approximation provides a useful description, but near a transition state, the approximation becomes increasingly inaccurate, as the spacing between energy levels decreases with increasing excitation.
A more realistic description of the inherently anharmonic potential, as also shown in Figure~\ref{fig:PES}, can be obtained by a Morse potential function~\cite{morse}, which leads to an energy expression of the form
\begin{equation}
    E_\text{vib}^\text{Morse}(i)= hc\omega_\text{e} \left( i + \frac{1}{2} \right)- hc\omega_\text{e} \chi_\text{e} \left( i + \frac{1}{2} \right)^2. \label{eq:energy-AHO}
\end{equation}
Here, the first term is equivalent to the harmonic potential, whereas anharmonicity is introduced in the second term.
With this, the spacing between the vibrational quantum levels is no longer uniform but decreases with increasing energy.
In the above equation, $\omega_\text{e}$ and $\omega_\text{e} \chi_\text{e}$ are spectroscopy constants, and $c$ is the speed of light.
The anharmonicity constant $\chi_\text{e}$ is related to the bond dissociation energy $D_\text{e}$ by
\begin{equation}
    \chi_\text{e} = \frac{hc\omega_\text{e}}{4D_\text{e}}.
\end{equation}
The fundamental frequencies and the anharmonicity constant can be obtained either from vibrational spectra or from ab-initio quantum calculations.
Although the Morse potential describes vibrational excitation and dissociation processes more accurately, it nonetheless is an approximation on the true multidimensional PES.

\subsection{Local modes}
\begin{figure}
    \centering
    \begin{subfigure}{0.4\textwidth}
    \centering



    \includegraphics{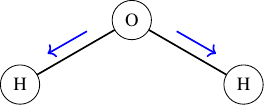}
    \caption{Symmetric stretch}
    \end{subfigure}
    \begin{subfigure}{0.4\textwidth}
    \centering



    \includegraphics{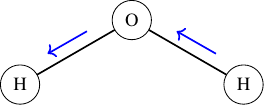}
    \caption{Asymmetric stretch}
    \end{subfigure}
    \caption{Vibrational modes of a water molecule in a normal mode basis.} \label{fig:normal}
\end{figure}

\begin{figure}
    \centering
    \begin{subfigure}{0.4\textwidth}
    \centering



    \includegraphics{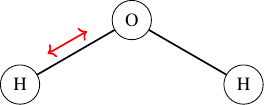}
    \caption{OH$_1$ stretch}
    \end{subfigure}
    \begin{subfigure}{0.4\textwidth}
    \centering



    \includegraphics{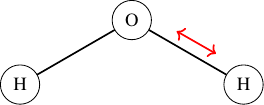}
    \caption{OH$_2$ stretch}
    \end{subfigure}
    \caption{Vibrational modes of a water molecule in a local mode basis.} \label{fig:local}
\end{figure}

So far, the discussion has been restricted to diatomic systems with a two-dimensional PES and a singular vibrational mode.
For polyatomic systems, the dimension of the PES increases to $3N-6$ non-zero degrees of freedom for non-linear molecules, or to $3N-5$ for linear molecules.
Commonly, molecular vibrations are described using normal mode analysis, in which the Hessian matrix evaluated at a stationary point on the PES is diagonalized to form a set of non-interacting harmonic oscillators.
The normal vibration modes represent a delocalized motion involving all nuclei of the molecule~\cite{lemus1999general}.
Although normal mode analysis is predominantly employed in the discussion of equilibrium geometries, its applicability becomes limited for highly excited vibrational states and for the description of localized bond-breaking processes in polyatomic systems.
Furthermore, the normal mode description arises from the harmonic approximation and might therefore be less appropriate for higher energies.
For an accurate description of bond strength and bond-specific dynamics, localized vibrational coordinates are required, such that only diatomic and triatomic fragments of the molecule participate in the vibration.
Local modes can be obtained from harmonically based normal modes through a unitary transformation based on an adiabatic connection scheme~\cite{zou2012relating}, given that the PES or experimental spectra are available for evaluation~\cite{konkoli1998new1, konkoli1998new2, konkoli1998new3}. The difference between normal and local stretching modes for the example of a water molecule are shown in Figures~\ref{fig:normal} and~\ref{fig:local}.

In both descriptions, two stretching modes appear for the water molecule. However, for the normal mode bases, both modes involve the stretching of two bonds, while only one bond at a time is involved for the stretching in the local basis. In many applications of local modes, stretching vibrational modes are treated with a local description, while bending modes are still sufficiently described on a normal mode basis.
In this mixed local-normal system, all stretching modes are described by independent Morse oscillators, whereas the bending modes are treated by harmonic oscillators.
The force constants in this basis provide an exact measure of the relative intrinsic bond strength and are therefore well suited for the description of bonding and reaction channels in molecules~\cite{zou2012relating}.

\subsection{Kinetic rate theory}
The distinction between localized and delocalized vibrational motion becomes particularly important in the description of reaction dynamics, as reactions involve the rearrangement of nuclear positions through the breaking and formation of chemical bonds.
Dissociation for example can be mainly associated, with energy localized in the coordinate corresponding to the breaking bond.
In addition, local modes provide a framework for the investigation of relaxation processes, where localized vibrational excitation is redistributed across the molecule.

\subsubsection{Unimolecular RRKM theory}
\begin{figure}
    \centering







    \includegraphics{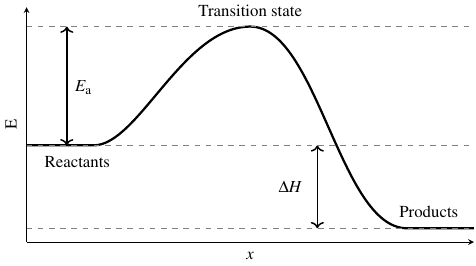}
    \caption{Energy profile along a reaction coordinate $x$ with activation energy $E_\text{a}$ and reaction enthalpy $\Delta H$.} \label{fig:rct}
\end{figure}

For unimolecular gas-phase reactions, the connection between the energetic state of the molecule and the chemical reaction rate is formalized in the Rice–-Ramsperger–-Kassel–-Marcus (RRKM) theory, which acts as a further extension of the Eyrings transition state theory~\cite{forst2012theory}.
In both theories, the energy profile along the reaction coordinate, as shown in Figure~\ref{fig:rct}, is the basis of the description.
According to both theories, provided the molecule has sufficient energy, it can overcome the reaction barrier $E_\text{a}$ and thus participate in a chemical reaction.
In contrast to Eyrings theory, the RRKM approach differentiates between a molecule with a given energy larger than the potential barrier and an activated complex, where most of the vibrational energy is localized in one or multiple reactive modes.
In the reactant molecule, the vibrational excitation is statistically distributed over a set of vibrational modes, which are loosely coupled, such that energy can flow between the intramolecular oscillator systems.
By this Intramolecular Vibrational Redistribution (IVR), energy is channeled into the reactive modes in between two collision events~\cite{nesbitt1996vibrational}.
In a purely harmonic system, vibrational modes are uncoupled and energy remains localized indefinitely, whereas a real molecular system exhibits anharmonic coupling effects that enable energy transfer between the individual modes.
For small systems, the vibrational energy periodically cycles through a limited number of coupled states on a short timescale compared to experimental resolution.
However, in larger molecules, energy redistribution typically occurs stepwise through networks of coupled modes.
The assumptions behind the RRKM theory hold whenever the IVR is sufficiently fast such that a statistical distribution of energy occurs.
Deviations may arise when vibrational energy remains uncoupled due to insufficient thermalization, where the energetic distribution in the molecule deviates from statistics due to a dynamic stabilization of near-resonance junctions, or in low-barrier conformational reactions~\cite{karmakar2020stable}.
For the system size considered in this work, IVR occurs on a much faster time scale than dissociation.
Mathematic models for the description of IVR have been developed for a set of triatomic systems, most notably by~\citet{bunker1962monte}.
In addition, empirical coupling models might be employed, such as the Harmonically--Coupled--Anharmonic--Oscillator (HCAO) approach~\cite{lemus1999general}, where anharmonic stretching modes are loosely coupled through harmonic bending modes.
RRKM theory itself considers the rotational contribution to the overall reaction rate by the distinction between active and inactive rotation.
Although approaches are made for a vector projection based evaluation of rotation~\cite{song2023theoretical}, no global criterion has yet been established to distinguish between active and inactive rotational modes.

\subsubsection{Bimolecular reactions and recombination} \label{sec:intro-bimolecular}
Bimolecular reactions in the RRKM theory occur whenever there is sufficient vibrational and translational motion to carry the reactants to the desired product state~\cite{crim1999vibrational}.
The reaction coordinate itself corresponds to a vibration with an imaginary frequency or a saddle point on the multidimensional PES.
For a reaction to occur, the system should retain an excitation along the reaction coordinate for a sufficiently long time in comparison to the reaction scale.
In contrast to unimolecular reactions, IVR occurs only close to the transition state.
Due to this much shorter interaction time, various reaction modes may have different efficiencies in promoting the reaction, and non-statistical behavior occurs.
This can be quantified using models such as the Sudden-Vector-Projection (SVP) approach, which measures the overlap between the vibrational mode vectors and the reaction coordinate~\cite{zhao2015state}.
Especially in larger polyatomic systems, many vibrational modes act merely as spectator modes.

Gas-phase recombination processes involve the stabilization of a transient collision complex through the collision with a third body, according to the Lindemann energy transfer (ET) or the Chaperon bound complex (BC) mechanism.
Both descriptions yield the same overall reaction rate~\cite{pack1972mechanism}.
As several kinetic details in both models remain generally unknown or require multiple assumptions, an alternative derivation of the recombination rate can be based on the equilibrium constant for the corresponding dissociation rate.

\section{The polyatomic HCAO model}
\begin{figure}
    \centering





    \includegraphics{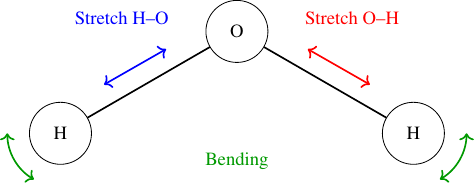}

    \caption{The H\textsubscript{2}O molecule in the local mode system with two stretching modes and one bending mode.}
    \label{fig:H2O}
\end{figure}

\begin{figure}
    \centering





    \includegraphics{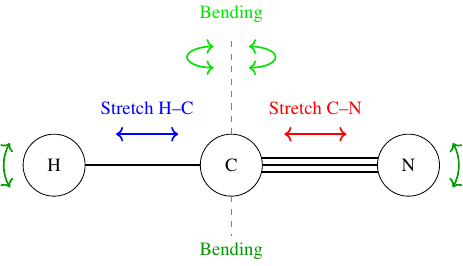}

    \caption{The HCN molecule in the local mode system with two stretching modes and two bending modes.}
    \label{fig:HCN}
\end{figure}

In this paper, the HCAO model is investigated in greater detail, where local stretching modes are divided into anharmonic stretching and harmonic bending modes.
As examples of the local mode system, Figures~\ref{fig:H2O} and~\ref{fig:HCN} show the H\textsubscript{2}O and HCN molecules with two local stretching modes each and one or two local bending modes defined.

All local stretching modes are modeled with the AHO model as described by~\citet{clement} for diatomic molecules.
Here, a Morse potential~\cite{morse} is used, where the energy levels are defined in Eq.~\ref{eq:energy-AHO} by solving the time-independent Schrödinger equation for a Morse potential.
In contrast, all bending modes are modeled as simple harmonic oscillators.
In addition to Eq.~\ref{eq:energy-SHO}, the vibrational energy can also be defined in terms of a characteristic temperature $\Theta_{\text{vib}}$ by
\begin{equation}
    E_{\text{vib}}^\text{HO}(i) = \left(i+\frac{1}{2}\right) k_\text{B} \Theta_{\text{vib}},
\end{equation}
where $k_\text{B}$ is the Boltzmann constant.
It is important to note that while harmonic vibrations in the local mode model may correspond to those in the normal mode description, this is not necessarily the case.

\subsection{Implementation of the polyatomic HCAO model}
The polyatomic HCAO model is implemented in the open-source particle code PICLas~\cite{piclas}.
To this point, only diatomic molecules can be modeled both with the vibrational anharmonic and harmonic oscillator models, whereas the vibrational modes in polyatomic models are treated harmonically in any case.
In the following, the changes required for the implementation of the HCAO model in PICLas are highlighted.

For the simulation of polyatomic molecules with the HCAO model, all vibrational energy levels and the spectroscopy constants for all local stretching modes are required.
For the bending modes, the characteristic temperature as harmonic parameter is a necessary input.

The vibrational temperature $T_{\text{vib},s}$ of a species $s$ can be found by a zero-point search over the mean vibrational energy, without considering the zero-point energy.
The vibrational energy of the stretching modes is based on the Boltzmann distribution of the population of the quantum levels.
\begin{alignat}{2}
	\left<E_{\text{vib},s}\right> &= &&\sum_{j_\text{bend}} E_{\text{vib},s,j_\text{bend}} + \sum_{j_\text{stretch}} E_{\text{vib},s,j_\text{stretch}} \nonumber \\
	&= &&\sum_{j_\text{bend}} \frac{k_\text{B} \Theta_{\text{vib},s,j_\text{bend}}}{\exp (\Theta_{\text{vib},s,j_\text{bend}} / T_{\text{vib},s}) - 1} \nonumber \\
	& &&+ \sum_{j_\text{stretch}} \left[ \sum_{i=0}^{i_\text{diss}} \left(E_{\text{vib},s,j_\text{stretch},i} - E_{\text{vib},s,j_\text{stretch},0} \right) \right. \nonumber \\
	& &&\left. \exp \left( - \frac{E_{\text{vib},s,j_\text{stretch},i} - E_{\text{vib},s,j_\text{stretch},0}}{k_\text{B} T_{\text{vib},s}}\right) \right] \nonumber \\
	& &&\left[ \sum_{i=0}^{i_\text{Diss}} \exp \left( - \frac{E_{\text{vib},s,j_\text{stretch},i} - E_{\text{vib},s,j_\text{stretch},0}}{k_\text{B} T_{\text{vib},s}}\right) \right]^{-1}.
\end{alignat}
With the vibrational temperature, the quantum numbers, the respective energy, and the vibrational degrees of freedom can now be calculated.

For the initialization process of the particles, either at the start of a simulation or if particles are entering the computational domain due to a surface flux, the quantum number $i_{j_\text{bend}}$ of each bending mode for a given vibrational temperature $T_{\text{vib},s}$ can be chosen using an equally distributed random number $r_j \in [0,1)$ per mode:
\begin{equation}
	i_{j_\text{bend}} = \left\lfloor \left(- \ln(r_j) \frac{T_{\text{vib},s}}{\Theta_{\text{vib},s,j_\text{bend}}}\right) \right\rfloor.
\end{equation}
For each stretching mode $j_\text{stretch}$ in the initialization process of particles, a quantum number $i_{j_\text{stretch}}$
\begin{equation}
	i_{j_\text{stretch}} = \left\lfloor \left( r_j i_\text{diss}\right) \right\rfloor + 1
\end{equation}
is chosen first, before deciding whether this quantum number is accepted according to the probability of this level.
$i_{j_\text{stretch}}$ is accepted if
\begin{equation}
	r_j < \frac{\exp \left( - E_{\text{vib},s,j_\text{stretch},i} / k_\text{B} T_{\text{vib},s} \right)}{\exp (- E_{\text{vib},s,j_\text{stretch},0} / k_\text{B} T_{\text{vib},s})}\label{eq:accept-quant}
\end{equation}
using a new random number $r_j$.

If a collision between two particles occurs, the post-collision quantum numbers are calculated based on the available collision energy $E_\text{coll}$, defined as the sum of the relative translational energy of the colliding particles $E_\text{rel}$, and the vibrational energy of the molecule undergoing relaxation $E_{\text{vib}}$
\begin{equation}
    E_\text{coll} = E_\text{rel} + E_{\text{vib}}.
\end{equation}
For the bending modes, the quantum number $i_{j_\text{bend}}$ is
\begin{align}
	i_{\text{max},j_\text{bend}} &= \left\lfloor \left(\frac{E_{\text{coll}}}{k_\text{B} \Theta_{\text{vib},s,j_\text{bend}}} - 0.5\right) \right\rfloor +1,\\
	i_{j_\text{bend}} &=\left\lfloor\left( r_j i_{\text{max},j_\text{bend}}\right) \right\rfloor
\end{align}
using a random number $r_j$.
The corresponding energy of the bending mode is subsequently defined as
\begin{equation}
	E_{\text{vib},s,j_\text{bend}} = \left(i_{j_\text{bend}} + \frac{1}{2}\right) k_\text{B} \Theta_{\text{vib},s,j_\text{bend}}.
\end{equation}
Considering the stretching modes, the maximum quantum number $i_{\text{max},j_\text{stretch}}$ for a given collision energy $E_{\text{coll}}$ is determined by identifying the highest quantum state for which the corresponding vibrational energy of the stretching mode remains below $E_{\text{coll}}$
\begin{equation}
	E_\text{coll} > E_{\text{vib},s,j_\text{stretch}}(i_{\text{max},j_\text{stretch}}).
\end{equation}
The quantum number is subsequently found drawing a random number $r_j$
\begin{equation}
	i_{j_\text{stretch}} = \left\lfloor\left( r_j i_{\text{max},j_\text{stretch}} \right) \right\rfloor.
\end{equation}
The corresponding energy of the stretching mode $E_{\text{vib},s,j_\text{stretch}}$ is then selected from the imported list of energy levels.
An acceptance-rejection method is used to determine whether the quantum numbers are accepted or the aforementioned process is repeated.
According to~\citet{paul-arm} using the Larsen--Borgnakke model~\cite{larsen-borgnakke}, the quantum numbers for all modes are accepted if a random number $r \in [0,1)$ is
\begin{equation}
    r < \frac{\left( E_\text{coll} - \sum_j E_{\text{vib},s,j}\right)^{\xi_\text{rel}/2-1}}{\left( E_\text{coll} - \sum_{j_\text{bend}} \frac{1}{2} k_\text{B} \Theta_{\text{vib},s,j_\text{bend}} - \sum_{j_\text{stretch}} E_{\text{vib},s,j_\text{stretch},0}\right)^{\xi_\text{rel}/2-1}}.
\end{equation}
$\xi_\text{rel}=5-2\omega_\text{ref}$ is the relative translational degree of freedom using the VHS model~\cite{bird} with the VHS viscosity parameter $\omega_\text{ref}$.
The process is repeated for a molecular collision partner.

While the vibrational degrees of freedom for the bending modes $j_\text{bend}$ are still being calculated following the harmonic approach, where the characteristic temperature of the mode may be used
\begin{equation}
	\xi_{\text{vib},s,j_\text{bend}} = \frac{2 E_{\text{vib},s,j_\text{bend}}}{k_\text{B} T_{\text{vib},s}} = \frac{2 \Theta_{\text{vib},s,j_\text{bend}} / T_{\text{vib},s}}{\exp\left(\Theta_{\text{vib},s,j_\text{bend}} / T_{\text{vib},s}\right) - 1},
\end{equation}
the ones for the stretching modes $j_\text{stretch}$ are now calculated with the energy of the mode that has been calculated as mentioned above:
\begin{equation}
	\xi_{\text{vib},s,j_\text{stretch}} = \frac{2 E_{\text{vib},s,j_\text{stretch}}}{k_\text{B} T_{\text{vib},s}}.
\end{equation}


\section{Polyatomic QK reaction modeling}
In PICLas, chemical reactions can either be modeled with the Total Collision Energy model~\cite{bird}, the
Quantum Kinetics (QK) model~\cite{bird-qk} or using cross sections.
The QK model is extended to different uni-, bi-, and trimolecular reactions including polyatomic molecules, which is presented in this section.
For both approaches, it first is to be determined if a reaction occurs.
Within the TCE model, this is done using reaction probabilities, based on the extended Arrhenius equation with the reaction rate coefficient $k(T)$
\begin{equation}
    k(T) = AT^b \exp \left(-\frac{E_{\mathrm{a}}}{k_{\mathrm{B}}T} \right).  \label{eq:arrhenius}
\end{equation}
The procedure for the QK model is explained in the following.

After determining that a reaction occurs, the remaining energy, i.e. the collision energy minus reaction enthalpy, is redistributed to the product species.
For this, the Larsen--Borgnakke model~\cite{larsen-borgnakke} is used, assuming that the the redistribution is equal over all available degrees of freedom according to the equipartition theorem.
For this, a root-finding algorithm is utilised to determine the vibrational degrees of freedom at a corresponding
collisional temperature~\cite{diss-paul}.

\subsection{Dissociation reactions}
Dissociation reactions, i.e. unimolecular reactions, are defined by
\begin{equation}
    A + M \rightleftharpoons B + C + M
\end{equation}
including an arbitrary collision partner $M$.
They may occur for each local stretching mode $j_\text{stretch}$ of a molecule $A$ if the available collision energy exceeds the dissociation energy of the respective mode.
The relevant collision energy is the relative translational energy of the colliding particles $E_{\text{tr},AM}$ plus the total vibrational energy of the dissociating polyatomic molecule $E_{\text{vib},A}$ minus the zero-point energies of all modes except the dissociating mode, which corresponds to the available energy.
Unlike the energies of all vibrational modes above ground state, their zero-point energies cannot be redistributed among the modes and are available only for the reaction of the local mode under consideration.
Additionally, the zero-point energies of the reaction products are subtracted since this amount of energy must remain after the reaction in order to form the products.
Finally, the collision energy is
\begin{align}
	E_{\text{coll},A,j_\text{stretch}} = \,\, & E_{\text{tr},AM} + E_{\text{vib},A} \nonumber \\
    &- \sum_{A,B,C} \left[ \sum_{j_\text{bend}} \frac{1}{2} k_\text{B} \Theta_{\text{vib},j_\text{bend}} + \sum_{j_\text{stretch}} E_{\text{vib},j_\text{stretch},0} \right] \nonumber \\
    &+ E_{\text{vib},A,j_\text{diss},0}. \label{eq:Ecoll-diss}
\end{align}
Subsequently, the excess energy is defined as the difference of the aforementioned collision energy and the dissociation energy of a stretching mode with
\begin{equation}
    E_{\text{excess},A,j_\text{stretch}} = E_{\text{coll},A,j_\text{stretch}} - D_{\text{e},A,j_\text{stretch}}.
\end{equation}
A dissociation may occur if
\begin{equation}
    E_{\text{excess},A,j_\text{stretch}} > 0.
\end{equation}


\subsection{Exchange reactions}
Exchange reactions are bimolecular reactions that are not necessarily directly coupled to vibrational stretching modes.
An example reaction is
\begin{equation}
    A + B \rightleftharpoons C + D.
\end{equation}
These reactions may occur in a first step, if the collision partner $B$ attacks the molecule $A$ in a specific region and angle, and in a second step, if the collision energy is sufficient to overcome the activation energy barrier of the reaction path.

An example for this steric dependence in the first step would be Cl attacking HCN to form HCl + CN.
This reaction can only occur, if the Cl radical approaches from the H-end.
To consider the dependence in the reaction model, a steric factor $\sigma \in [0,1]$ is defined as the probability that the collision event may occur in a way that leads to a reaction.
For the above example, $\sigma$ could be defined as 0.5.
A random number criterion $r \in [0,1) < \sigma$ is then employed to define if the exchange reaction pathway is sterically possible.

For the second step, a collision energy is first defined, which is the relative translational energy of the colliding particles $E_{\text{tr},AB}$ plus the vibrational energy of molecule $A$.
Due to the shorter interaction time of the molecules in comparison to a unimolecular reaction, not all vibrational modes may fully couple to the reactive mode.
Thus, a reaction-specific parameter $\alpha_j \in [0,1]$ defines the degree of coupling of each vibrational mode $j$.
This is further defined, for example, in the SVP model described in Section~\ref{sec:intro-bimolecular}.
For the example of a methane exchange reaction, as discussed in Section~\ref{ssec:ch4}, classical trajectory simulations showed that $\alpha_j=0.21$ for this pathway~\cite{jordan1995classical}.
Finally, the collision energy is
\begin{equation}
    E_{\text{coll},AB} = E_{\text{tr},AB} + \sum_{j} \alpha_{j} E_{\text{vib},A,j}. \label{eq:Ecoll-exch}
\end{equation}
Following this, the excess energy can be defined as the difference between the aforementioned collision energy and the activation energy with
\begin{equation}
    E_{\text{excess},AB} = E_{\text{coll},AB} - E_{\text{a},AB}.
\end{equation}
Note that the activation energy $E_{\text{a},AB}$ is not the same as for Arrhenius reactions.
An exchange reaction may occur if
\begin{equation}
    E_{\text{excess},AB} > 0.
\end{equation}

\subsection{Recombination reactions} \label{ssec:rec}
A recombination reaction is the backward reaction of a dissociation reaction defined as
\begin{equation}
    B + C + M \rightleftharpoons A + M.
\end{equation}
The reaction rate is calculated using the forward reaction rate and the equilibrium constant with partition sums~\cite{bird-qk}.
The forward rate is defined as
\begin{equation}
    k_\text{f} = P \frac{2 \left(d_\text{ref}\right)^2}{\delta_\text{sym}} \sqrt{\frac{2 \pi k_\text{B} T_\text{ref}}{m_\text{red}}} \left(\frac{T}{T_\text{ref}}\right)^{1-\omega_\text{ref}},
\end{equation}
where $d_\text{ref}$ is the VHS reference diameter defined at a VHS reference temperature $T_\text{ref}$, $\omega_\text{ref}$ again is the VHS viscosity parameter, $\delta_\text{sym}$ is a symmetry factor equal to two for similar species and one for dissimilar species, and $m_\text{red}$ is the reduced mass.
The VHS parameters are found as mean values from both the involved species $A$ and $M$.
If direct VHS data are available for each reaction combination, these can be used accordingly.
For the probability $P$ of the forward reaction at a given temperature, all possible combinations of individual vibrational levels are evaluated and their contribution weighted by the use of partition sums.
If the energy of the vibrational states, under consideration of the translational energy, exceeds the dissociation energy, the reaction may take place:
\begin{widetext}
\begin{equation}
    P = \frac{\sum_{j_1=1}^{N_1} \sum_{j_2=1}^{N_2} \sum_{j_3=1}^{N_3} \dots \left[ \Gamma_\text{inc} \left(5/2-\omega_\text{ref}; \frac{D_\text{e}-\left(\epsilon_{j_1}+\epsilon_{j_2}+\epsilon_{j_3}+\dots\right)}{k_\text{B} T}\right) \exp\left(\frac{-\left(\epsilon_{j_1}+\epsilon_{j_2}+\epsilon_{j_3}+\dots\right)}{k_\text{B} T}\right) \right]}{\sum_{j_1=1}^{N_1} \sum_{j_2=1}^{N_2} \sum_{j_3=1}^{N_3} \exp\left(\frac{-\left(\epsilon_{j_1}+\epsilon_{j_2}+\epsilon_{j_3}+\dots\right)}{k_\text{B} T}\right)}.\label{eq:recomb-correct}
\end{equation}
\end{widetext}
As the computational effort increases fast with the number of vibrational modes, an approximation of this is done using evenly spaced sample energy levels between the ground state and the dissociation level:
\begin{equation}
    P = \frac{\sum_{i=1}^{N_\text{samp}} \Gamma_\text{inc} \left(5/2-\omega_\text{ref}; \frac{D_\text{e}-E_i}{k_\text{B} T}\right) \rho(E_i) \exp\left(\frac{-E_i}{k_\text{B} T}\right)}{\sum_{i=1}^{N_\text{samp}} \rho(E_i)\exp\left(\frac{-E_i}{k_\text{B} T}\right)}.
\end{equation}
The newly introduced density of vibrational energy states $\rho(E_i)$ is calculated using the Whitten--Rabinovich (WR) expression~\cite{whitten-rabinovich,stace-sellers,bao-truhlar}, based on a molecule consisting of $l$ harmonic oscillators
\begin{equation}
    \rho(E_i) = \frac{\left[ E + a(E) E_\text{z} \right]^{l-1}}{(l-1)! \prod_{k=1}^l h \nu_k},
\end{equation}
where
\begin{alignat}{2}
    a(E) &= 1 - \beta \Omega, && \\
    \log \Omega & = -1.0506 (E/E_\text{z})^{0.25}, \quad &&\text{when } E/E_\text{z} \geq 1, \\
    \Omega^{-1} &= 5(E/E_\text{z}) + 2.73(E/E_\text{z})^{1/2} + 3.51 , \quad &&\text{when } E/E_\text{z} < 1, \\
    \beta &= \frac{(l-1)^2 \sum_{k=1}^l \nu_k^2}{l\left(\sum_{k=1}^l \nu_k\right)^2},
\end{alignat}
with vibrational frequency $\nu_k$ of oscillator $k$, zero-point energy $E_\text{z}$, and vibrational energy $E$ above ground level.
The vibrational frequencies are calculated from the harmonic or anharmonic models for the bending and stretching modes, respectively.
Due to the use of anharmonic fundamental frequencies, the expression already contains some anharmonicity.
However, the approach of the WR approximation using harmonic oscillators leads to a small deviation compared to the correct but computationally expensive approach using Eq.~\ref{eq:recomb-correct}.
Still, to the authors knowledge, the WR approximation is the only feasible approach without significantly larger computational costs.
Further anharmonic correction factors could be investigated in the future.

\subsection{Rotation coupling}
If a rotational-vibrational coupling is present, the equations for the collision energies need to be adapted as follows with a degree of coupling $\gamma \in [0,1]$.
The degree of coupling may for example be based on the SVP approach, as stated in Section~\ref{sec:intro-bimolecular}.
The collision energy for the dissociation reaction (Eq.~\ref{eq:Ecoll-diss}) is extended to
\begin{align}
	E_{\text{coll},A,j_\text{stretch}} = \,\, & E_{\text{tr},AM} + \gamma E_{\text{rot},A} + E_{\text{vib},A} \nonumber \\
    &- \sum_{A,B,C} \left[ \sum_{j_\text{bend}} \frac{1}{2} k_\text{B} \Theta_{\text{vib},j_\text{bend}} + \sum_{j_\text{stretch}} E_{\text{vib},j_\text{stretch},0} \right] \nonumber \\
    &+ E_{\text{vib},A,j_\text{diss},0}.
\end{align}
For the exchange reaction, the collision energy defined in Eq.~\ref{eq:Ecoll-exch} changes to
\begin{equation}
    E_{\text{coll},AB} = E_{\text{tr},AB} + \gamma E_{\text{rot},A} + \sum_{j} \alpha_{j} E_{\text{vib},A,j}.
\end{equation}

\subsection{Competing reactions}
If both dissociation and exchange reactions are possible (i.e. more than one reaction channel has an excess energy with a positive value), a random number $r \in [0,1)$ is drawn to decide which reaction occurs.
This decision is weighted according to the amount of excess energy $E_{\text{excess}}$ of each available reaction path.

\section{Verification}
To verify the newly developed HCAO-based reaction model, a series of closed-system reservoir simulations involving different molecular species is performed.
In all simulations, the computational domain consists of a single reservoir cell with a volume of $V = \num{4.64E-6}$ m$^3$.
Consequently, the observed species evolution is solely governed by chemical reactions and relaxation processes.
Furthermore, all simulation cases with the different reaction models employed the HCAO-based vibrational model to minimize deviations due to anharmonic energy effects.
For the investigated reaction systems, calculations either directly showed that rotational effects do not contribute to the reaction~\cite{jordan1995classical} or no indication for significant rotational contribution could be found in the literature.
Therefore, these effects were not taken into account in the following.

\subsection{Water}
\begin{figure}
    \centering
    \includegraphics{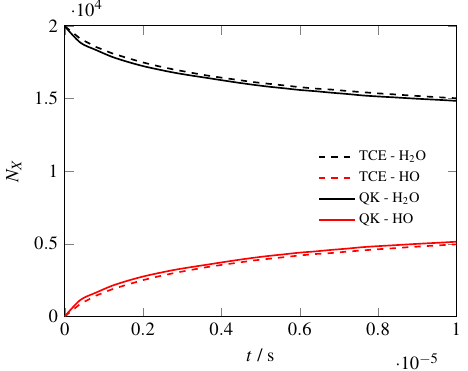}
    \caption{Temporal evolution of the simulation particle number for the dissociation of water. Compared are DSMC simulations using the Arrhenius-based TCE model and the QK model in combination with the HCAO approach. \label{fig:reac_h2o}}
\end{figure}

\begin{figure}
    \centering
    \includegraphics{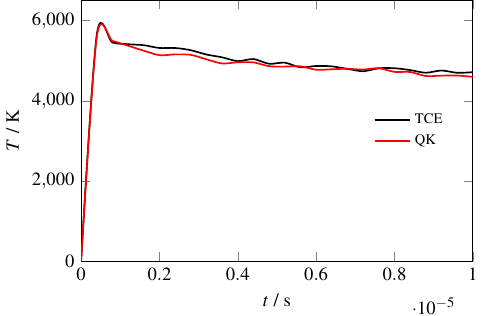}
    \caption{Temporal evolution of the OH temperature during the dissociation of water. Compared are DSMC simulations using the Arrhenius-based TCE model and the QK model in combination with the HCAO approach. \label{fig:temp_h2o}}
\end{figure}

As a first validation case, the dissociation of water under the formation of a hydrogen atom and a hydroxyl radical is investigated:
\begin{equation}
    \mathrm{H_2O} + \mathrm{N_2} \rightarrow  \mathrm{H} + \mathrm{OH} + \mathrm{N_2}.
\end{equation}
Due to the symmetry of the water molecule, the dissociation of either OH stretching bond results in identical product species.
The simulation is initialized at a temperature $T_0 =$ 6000~K, with an initial water number density $n_{0,\mathrm{H_2O}}$ = \num{1E+23}~m$^{-3}$.
Furthermore, molecular nitrogen is introduced as a non-reactive collision partner at a concentration $n_{0,\mathrm{N_2}}$ = \num{1E+23}~m$^{-3}$.
For the water molecule, the spectroscopic constants are summarized in Table~\ref{tab:h2o}.
\begin{table}\renewcommand{\arraystretch}{1.4}
\centering
\caption{Spectroscopic constants for the different vibrational modes of water, according to the HCAO model \cite{halonen1988fermi, wright1985potential}. \label{tab:h2o}}
\begin{tabular}{lll}
\hline
Mode                  & 1      & 2,3 \\ \hline
Stretch               & -      & HO \\
$\Theta_{\text{vib}}$ / K & 2294.3 & \\
$\omega_\text{e}$ / m$^{-1}$    &        & 387620.0 \\
$\chi_\text{e}$ &        & 0.021774 \\ \hline
\end{tabular}
\end{table}
For the diatomic species OH and N$_2$, a single Morse potential function is used to describe the vibrational energy levels.
As the bonding characteristics of the OH radical are similar to those of the OH bond in water, the same Morse potential parameters are employed in both cases for the stretching mode.
For N$_2$, the parameters $\omega_\text{e} = \num{234599.0}$~m$^{-1}$ and $\chi_\text{e}=\num{0.007439}$ are chosen.
To evaluate the accuracy of the implemented reaction model, the HCAO-based approach is compared to the established TCE model.
The Arrhenius parameters used in this reference simulation are summarized in Table~\ref{tab:h2o}.
\begin{table}\renewcommand{\arraystretch}{1.4}
\centering
\caption{Arrhenius parameter for the dissociation of water. \label{tab:h2o}}
\begin{tabular}{lll}
\hline
$A$ / m$^3$ s$^{-1}$ K$^{-b}$ & $b$           & $E_{\text{a}}$ / K \\ \hline
       \num{8E-15}                       & 0.0 & 52920                            \\ \hline
\end{tabular}
\end{table}
For all simulations, a uniform particle weight of $w=\num{5E2}$ and a timestep $\Delta t =\num{2E-9}$~s are employed.
In a first instance, the temporal evolution of the species particle number is evaluated in Figure~\ref{fig:reac_h2o}.
As the simulation time progresses, the concentration of H$_2$O decreases, while the product OH concentration increases correspondingly.
With the newly developed polyatomic reaction model, the concentration change predicted by the Arrhenius-based reference simulation is reproduced with very good accuracy.
In addition to the species concentrations, the total temperature of the product species OH is investigated in Figure~\ref{fig:temp_h2o}.
An initially rapid increase in the temperature up to 6000~K can be observed, as the first reaction products are formed.
Over the course of the simulation, the OH radicals relax during collisions with the surrounding gas mixture until a value of approximately 4700~K is reached.
The temperature evolution predicted by the HCAO-based QK model is in very good agreement to the simulation with the reference TCE simulation.
Thus, not only the reaction rates but also the energy of the product species is accurately reproduced.

\subsection{Hydrogen cyanide}
\begin{figure}
    \centering
    \includegraphics{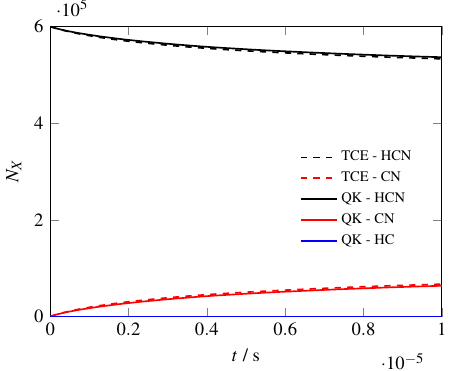}
    \caption{Temporal evolution of the simulation particle number for the dissociation of hydrogen cyanide at 5000~K. Compared are DSMC simulations using the Arrhenius-based TCE model and the QK model in combination with the HCAO approach.} \label{fig:hcn1}
\end{figure}

\begin{figure}
    \centering
    \includegraphics{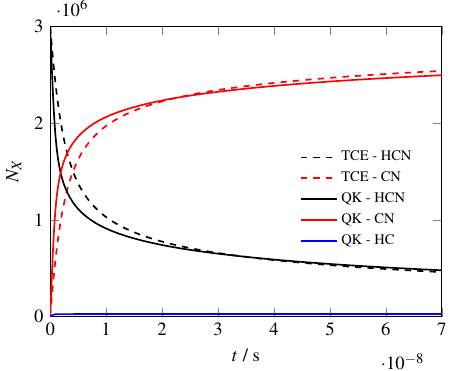}
    \caption{Temporal evolution of the simulation particle number for the dissociation of hydrogen cyanide at 12000~K. Compared are DSMC simulations using the Arrhenius-based TCE model and the QK model in combination with the HCAO approach.} \label{fig:hcn2}
\end{figure}

As a second validation case, the two possible dissociation channels of hydrogen cyanide are investigated:
\begin{equation}
    \mathrm{HCN} + \mathrm{N} \rightarrow \mathrm{H} + \mathrm{CN} + \mathrm{N},     \label{eq:rct_cn}
\end{equation}
\begin{equation}
    \mathrm{HCN} + \mathrm{N} \rightarrow \mathrm{HC} + \mathrm{2N}.    \label{eq:rct_ch}
\end{equation}
HCN is an asymmetric linear molecule, for which two different product species might be formed.
However, due to the significantly higher bond strength of the CN triple bond compared to the HC single bond, the dissociation channel leading to the formation of the cyanide radical is energetically favored.
As in the previous test case, first an Arrhenius-based simulation is employed and subsequently used as a reference for the evaluation of the HCAO-based QK reaction model.
Meaningful reaction kinetics according to the Arrhenius equation could only be found for the dissociation under the formation of CN.
Therefore, only the dominant reaction pathway was considered for the Arrhenius-based model, while both dissociation channels were included in the HCAO-based simulation.
The simulations are initialized at two temperatures $T_{0,1} =$ 5000~K and $T_{0,2} =$ 12000~K, with both hydrogen cyanide and the non-reacting collision partner atomic nitrogen at a number density $n_{0,\mathrm{HCN}}$ = $n_{0,\mathrm{N}}$ = \num{3E+24}~m$^{-3}$.
The spectroscopic constants for HCN are summarized in Table~\ref{tab:hcn}.
\begin{table}\renewcommand{\arraystretch}{1.4}
\centering
\caption{Spectroscopic constants for the different vibrational modes of hydrogen cyanide, according to the HCAO model \cite{halonen1988fermi, wright1985potential}. \label{tab:hcn}}
\begin{tabular}{llll}
\hline
Mode                  & 1,2      & 3      & 4       \\ \hline
Stretch               & -      & HC     & CN      \\
$\Theta_{\text{vib}}$ / K & 1023.9 &        &         \\
$\omega_\text{e}$ / m$^{-1}$     &        & 342040 & 211920  \\
$\chi_\text{e}$       &        & 0.0146 & 0.00368 \\ \hline
\end{tabular}
\end{table}
For the product species CN, a single Morse potential function is employed, with $\omega_\text{e} = \num{211920}$~m$^{-1}$ and $\chi_\text{e}=\num{0.0146}$.
The spectroscopic parameter $\omega_\text{e} = \num{342040}$~m$^{-1}$ and $\chi_\text{e}=\num{0.0146}$ are used to model the vibrational states of the HC product.
All vibrational parameter are based on Ref.~\cite{zheng2003coset}.
The Arrhenius reaction parameters for the dissociation of the HC bond are summarized in Table~\ref{tab:hcn_tce}.
\begin{table}\renewcommand{\arraystretch}{1.4}
\centering
\caption{Arrhenius parameter for the dissociation of hydrogen cyanide. \label{tab:hcn_tce}}
\begin{tabular}{lll}
\hline
$A$ / m$^3$ s$^{-1}$ K$^{-b}$ & $b$           & $E_{\text{a}}$ / K \\ \hline
       \num{5.9283E-04}                       & -2.6 & 62845                            \\ \hline
\end{tabular}
\end{table}
A particle weight of $w=100$ and a timestep $\Delta t = \num{1E-10}$~s are chosen for all simulations.
In Figure~\ref{fig:hcn1}, the temporal evolution of the species particle numbers obtained with the different reaction models are compared for the temperature $T_{0,1} =$ 5000~K.
At this temperature, the dissociation of hydrogen cyanide under formation of CN progresses with a comparably slow rate.
In both simulation cases, no HC molecule is formed, as no sufficiently energetic reactant molecules are present, which can overcome the dissociation barrier of the triple bond.
The different reaction models predict a temporal evolution of the species particle number in very good agreement with each other.
In Figure~\ref{fig:hcn2}, the particle number profiles for the different reaction models at a temperature $T_{0,1} =$ 12000~K are shown.
At the start of the simulation, the HCAO-based QK model predicts a higher dissociation rate compared to the Arrhenius-based reference simulation.
This deviation might be attributed to the very high temperatures in the investigated test case, which significantly exceeds the considered range in the calculation of the Arrhenius reaction parameters~\cite{tsang1991chemical}.
Therefore, the TCE reaction model might no longer fully capture the inherent reaction dynamics.
In addition, at this higher temperature, the HCAO-based model also predicts a minor contribution from the dissociation pathway according to Eq.~\ref{eq:rct_ch}, in which a small fraction of hydrogen cyanide molecules dissociate through the formation of HC radicals.
As the simulation progresses, both approaches converge toward a similar equilibrium concentration.
The remaining differences in the simulated particle numbers are mainly caused by the inclusion of the two competing dissociation channels in the HCAO-based model.
While the reference simulation considers only the dominant reaction pathway, the HCAO-based QK model directly accounts for additional mechanisms, which could occur simultaneously in a real gas system.
A small fraction of highly-energetic molecules may overcome the dissociation barrier for the triple bond and enable the formation of HC radicals.
Therefore, neglecting such pathways may lead to an incomplete description for high-energetic and high-temperature systems.
Overall, the HCAO-based model allows to accurately capture different dissociation processes, without requiring prior knowledge of all individual kinetic rate coefficients.

\subsection{Methane}\label{ssec:ch4}
\begin{figure}
    \centering
    \includegraphics{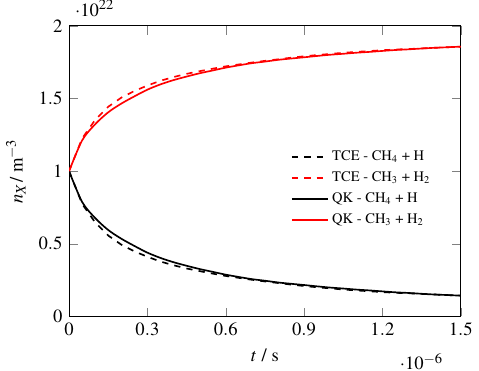}
    \caption{Temporal evolution of the species number densities for the CH$_4$ + H $\rightarrow$ CH$_3$ + H$_2$ exchange reaction. Compared are DSMC simulations using the Arrhenius-based TCE model and the QK model in combination with the HCAO approach. } \label{fig:ch4}
\end{figure}

To verify the reaction model for bimolecular systems and recombination, different reaction channels of methane are simulated.
In a first instance, the hydrogen exchange reaction of methane is investigated:
\begin{equation}
    \mathrm{CH_4} + \mathrm{H} \rightarrow \mathrm{CH_3} + \mathrm{H_2}.
\end{equation}
In this exchange reaction, two polyatomic species appear.
For methane, nine different vibrational modes need to be considered, while the methyl radical employs six vibrational modes.
The simulation case is initialized at a temperature $T_0 =$ 5000~K, with the reactant and the product species at a number density $n_{0,X}$ = \num{1E+22}~m$^{-3}$.
The spectroscopic constants for methane are summarized in Table~\ref{tab:ch4} and those for the methyl radical are summarized in Table~\ref{tab:ch3}.
\begin{table}\renewcommand{\arraystretch}{1.4}
\centering
\caption{Spectroscopic constants for the different vibrational modes of methane, according to the HCAO model \cite{halonen1997internal}. \label{tab:ch4}}
\begin{tabular}{llll}
\hline
Mode                  & 1,2      & 3,4,5      & 6,7,8,9 \\ \hline
Stretch               & -      & -      & CH \\
$\Theta_{\text{vib}}$ / K & 1565.2 & 1360.3 & \\
$\omega_\text{e}$ / m$^{-1}$     &        &        & 312350 \\
$\chi_\text{e}$       &        &        & 0.0190 \\ \hline
\end{tabular}
\end{table}
\begin{table}\renewcommand{\arraystretch}{1.4}
\centering
\caption{Spectroscopic constants for the different vibrational modes of CH$_3$, according to the HCAO model \cite{westre1991force}. \label{tab:ch3}}
\begin{tabular}{llll}
\hline
Mode                  & 1     & 2,3      & 4,5,6 \\ \hline
Stretch               & -     & -      & CH \\
$\Theta_{\text{vib}}$ / K & 872.1 & 2016.2 & \\
$\omega_\text{e}$ / m$^{-1}$    &       &        &  312350 \\
$\chi_\text{e}$       &       &        & 0.0190 \\ \hline
\end{tabular}
\end{table}
As for the water molecule, the bonding situation in the dissociated state does not significantly differ from the reactant state.
For molecular hydrogen, a single Morse potential function is employed, with $\omega_\text{e} = \num{439524}$~m$^{-1}$ and $\chi_\text{e}=\num{0.0287}$.
For comparison, a reference Arrhenius-based simulation is performed, with the reaction parameter summarized in Table~\ref{tab:ch4_tce}.
\begin{table}\renewcommand{\arraystretch}{1.4}
\centering
\caption{Arrhenius parameter for the hydrogen exchange reaction. \label{tab:ch4_tce}}
\begin{tabular}{lll}
\hline
$A$ / m$^3$ s$^{-1}$ K$^{-b}$ & $b$           & $E_{\text{a}}$ / K \\ \hline
       \num{1.03E-24}& 2.5& 4825\\ \hline
\end{tabular}
\end{table}
Contrary to the dissociation treatment in the new reaction model, for which all necessary information is contained in the vibrational modes, bimolecular reactions require the specification of an activation energy.
For the exchange reaction considered here, an energetic barrier of \num{1.03E-19}~J is employed~\cite{jordan1995classical}.
Furthermore, the energy of the various vibrational modes of methane does not fully contribute to the reaction.
Rather, only approximately 21~\% of the energy couples to the reaction mode and is therefore available to overcome the reaction barrier~\cite{jordan1995classical}.
In all simulations, a particle weight of $w=100$ and a timestep $\Delta t = \num{5E-9}$~s is used.
The temporal evolution of the species concentration with the different reaction models is evaluated in Figure~\ref{fig:ch4}.
At the beginning of the simulation, both reactant and product species are present at equal concentrations.
As the simulation progresses, the reactants are rapidly consumed, until an equilibrium value is reached.
Again, both the HCAO-based QK model and the TCE reaction model are in very good agreement with other.

\begin{figure}
    \centering
    \includegraphics{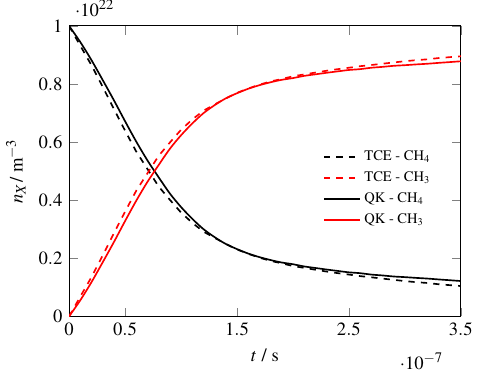}
    \caption{Temporal evolution of the species number densities for the dissociation and recombination equilibrium of methane. Compared are DSMC simulations using the Arrhenius-based TCE model and the QK model in combination with the HCAO approach. } \label{fig:ch4_diss}
\end{figure}

In a second step, the dissociation and recombination equilibrium of methane is investigated:
\begin{equation}
    \mathrm{CH_4} + \mathrm{M} \rightleftharpoons \mathrm{CH_3} + \mathrm{H} + \mathrm{M}.
\end{equation}
The collision partner M can be any species present in the system, i.e., CH$_4$, CH$_3$, or H.
For methane and the methyl radical, again the spectroscopic constants as detailed in Table~\ref{tab:ch4} and~\ref{tab:ch3} are employed.
In both the Arrhenius reference simulation and the QK simulation case, the forward dissociation rate is modeled directly, while the backward recombination rate is calculated from the equilibrium constant and the partition functions, as detailed in Section~\ref{ssec:rec}.
The Arrhenius constants for the dissociation are summarized in Table~\ref{tab:ch4_diss}.
\begin{table}\renewcommand{\arraystretch}{1.4}
\centering
\caption{Arrhenius parameter for the dissociation of methane. \label{tab:ch4_diss}}
\begin{tabular}{lll}
\hline
$A$ / m$^3$ s$^{-1}$ K$^{-b}$ & $b$           & $E_{\text{a}}$ / K \\ \hline
    \num{5.26E-12}   & 0.0 & 51488 \\ \hline
\end{tabular}
\end{table}
A temperature $T_0 =$ 9000~K is chosen in the simulation setup, with all species at a number density $n_{0,X}$ = \num{1e+23}~m$^{-3}$.
In Figure~\ref{fig:ch4_diss}, the temporal evolution of the species concentration is compared for the different reaction models.
Even though both the recombination and dissociation reaction can take place, the dissociation reaction proceeds with a higher rate, such that the methane concentration decreases over time.
Both the Arrhenius- and the HCAO-based reaction model are in good agreement with each other.
Thus, the newly developed reaction model does not only reproduce unimolecular dissociation processes with high accuracy, but also bimolecular reactions and recombination mechanisms.

\section{Conclusion}
A novel reaction model for polyatomic species in the Direct Simulation Monte Carlo (DSMC) method has been developed, which allows to accurately represent anharmonic vibrational effects and their influence on chemical reactions.
The model makes use of a mixed local-normal mode description for the vibrational relaxation based on the HCAO concept.
The anharmonic reaction modeling is based on RRKM theory, where the reaction probability and corresponding rate coefficients are determined from the available vibrational and translational energy as well as the coupling fraction between internal energy modes.
This approach allows to simulate energy-dependent reaction rates without requiring a fit to Arrhenius parameters.
All gas-phase reaction processes as typically encountered in DSMC applications can be investigated.
The accuracy of the developed reaction model has been evaluated for several representative molecular systems.
For all investigated cases, the simulations were in very good agreement with the established Arrhenius-based reaction model.

In future work, the model could be further improved by an extended treatment of anharmonic effects in recombination processes.
The model could be validated by the comparison with experimental measurements, especially to quantify the influence of anharmonicity under extreme thermochemical conditions, such as atmospheric re-entry flows.
Also, a further extension to state-specific reaction modeling could be envisioned, as done by~\citet{kustova}.

\begin{acknowledgments}
This work is funded by the Deutsche Forschungsgemeinschaft (DFG, German Research Foundation) – Project-ID 516238647 – SFB 1667/1 (ATLAS - Advancing Technologies of Very Low-Altitude Satellites).
\end{acknowledgments}

\section*{Data Availability Statement}
The data that support the findings of this study are available from the corresponding author upon reasonable request.

\nocite{*}
\bibliography{bib}

\end{document}